\documentclass{article}

\usepackage{graphicx}

\def\giorno{5/3/2020}

\def\a{\alpha}

\def\ga{\gamma}
\def\s{\sigma}

\def\^#1{\widehat{#1}}

\def\beql#1{\begin{equation} \label{#1}}
\def\beq{\begin{equation}}
\def\eeq{\end{equation}}

\def\<{\langle}
\def\>{\rangle}

\def\eqref#1{(\ref{#1})}

\begin{document}

\title{Data analysis for the COVID-19 early dynamics in Northern Italy}

\author{G. Gaeta \\ {\it Dipartimento di Matematica, Universita' degli Studi di Milano} \\ {\it via Saldini 50, I-20133 Milano (Italy)} \\ and \\ {\it SMRI, 00058 Santa Marinella (Italy)} \\ {\tt giuseppe.gaeta@unimi.it} }

\date{\giorno}

\maketitle

The COVID-19 epidemics, started in China in January 2020\footnote{Data analysis from various sources, see e.g. \cite{ICR}, suggest first cases may have developed earlier on, albeit not recognized as due to a new virus.}, was recognized to have reached Italy around February 20; recent estimates show that most probably the virus circulated in the country already in January, but was not recognized.

The development of the epidemic in Italy was characterized by a rather large number of new cases even in early days. This led to the isolation of two relatively small areas -- one with about 50,000 inhabitants and the other with bout 3,000 ones -- and in these areas the search for infected cases has been performed by a wide number of laboratory exams, which led to the discovery of a rather large number of cases with weak or no symptoms at all. In fact,about 50\% of identified contagion cases are being teated by simple home isolation.

This should be compared with early data from China and in particular in Wuhan (where the huge number of both serious medical cases and the total population presumably prevented from screening contacts of infected people not showing symptoms, except maybe for medical doctors most involved in fighting the virus), which reported a very small number, of the order of a few percent, of asymptomatic contagion \cite{CDC}.

On the other hand, the Chinese report made a well defined difference between serious and less seriously affected patients, showing that the first case amounted to about 20\% of the total COVID-19 patients, all of whom were hospitalized.

In this note we want to discuss raw data for the Italian situation; in a companion paper we will put forward a very simple model (of the SIR type \cite{Murray}) taking into account the presence of a large number of asymptomatic cases, many of which are most probably not detected nor detectable except in very small communities\footnote{We understand that in the smaller of the isolated communities in Northern Italy, totalizing about 3,000 inhabitants, a mass screening is being conducted; this should provide more reliable data on the fraction of asymptomatic infections, together with other precious information.}; this will then be used to understand how much a prompt identification of similar cases (not practically feasible at this stage, but which could become possible if rapid and economic tests would be commercially developed) could help in reducing the spread of the infection.

The used data were obtained from the publicly available ones provided by the daily Italian ``Protezione Civile'' reports \cite{PCrep} ; the total number of cases in the country are also reported by W.H.O. \cite{WHOrep}. General facts about the COVID-19 epidemics in China can be accessed though the China CDC early report \cite{CDC} and through the WHO-CDC document \cite{WHOCDC}; statistical and philogenetic analyses are made publicby the Imperial College group \cite{ICR}.

\section{Raw data}

\subsection{Benchmark: China}

\begin{figure}
  \includegraphics[width=250pt]{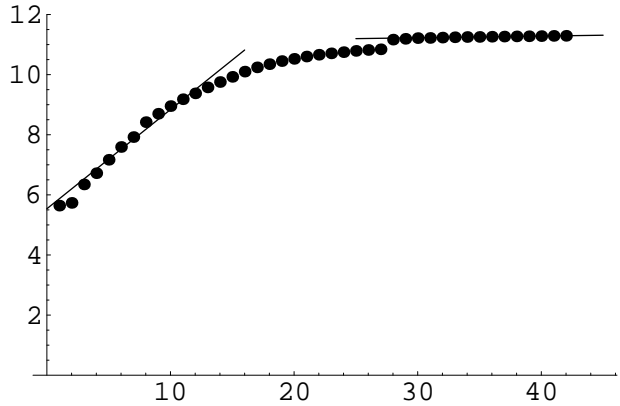}
  \\
  \caption{Epidemiological data for all of China,as provided by WHO \cite{WHOrep}. Cumulative cases in China; these are fitted by exponential laws \eqref{eq:alpha} corresponding to best fits for days 3-13 and for days 32-42.}\label{fig:china}
\end{figure}

We start by recalling some facts and figures about the situation in China.
The spreading of COVID-19 in all of China (thus not making a distinction between the Hubei region and the rest of the country) is summarized in the semi-logarithmic plot of Figure \ref{fig:china}. This shows that the growth rate $\a$ for the number of total cases in day $k$,
\beql{eq:alpha} n(k) \ = \ \exp[ \a \,k ] \ n(0) \eeq
passed from $\a_i = 0.330 $ for the early period (in particular, this fit was obtained for the period January 23 -- February 2) to $\a_f = 0.006 $ for the last available data (in particular, this fit was obtained for for the period February 21 -- March 2), showing that the drastic measures taken by the Chinese government produced a substantial effect.

The doubling time $\tau$, obtained from the above via
\beql{eq:tau} \tau \ = \ \a^{-1} \ \log (2) \ , \eeq
passed correspondingly from $\tau_i=2.1$ to $\tau_f = 122$.

The daily growth factor $\gamma$ is determined as
\beql{eq:gamma} \ga \ = \ \exp [ \a ] \ ; \eeq
this passed from $\ga_i = 1.391$ to $\ga_f = 1.006$.

\begin{center}

{\small

\begin{tabular}{|r|r||r|r||r|r||r|r||r|r||}
\hline
day & cases & day & cases & day & cases & day & cases & day & cases \\
\hline
1 &  282 & 2 & 309 & 3 & 571 & 4 & 830 & 5 & 1297 \\
6 & 1985 & 7 & 2761 & 8 & 4537 & 9 & 5997 & 10 & 7736 \\
11 & 9720 & 12 & 11821 & 13 & 14411 & 14 & 17238 & 15 & 20471 \\
16 & 24363 & 17 & 28060 & 18 & 31211 & 19 & 34598 & 20 & 37251 \\
21 & 40235 & 22 & 42708 & 23 & 44730 & 24 & 46550 & 25 & 48548 \\
26 & 50054 & 27 & 51174 & 28 & 70635 & 29 & 72528 & 30 & 74280 \\
31 & 74675 & 32 & 75569 & 33 & 76392 & 34 & 77042 & 35 & 77262 \\
36 & 77780 & 37 & 78191 & 38 & 78630 & 39 & 78961 & 40 & 79394 \\
41 & 79968 & 42 & 80174 &    &       &    &       &    &       \\
\hline
\end{tabular}

}

\medskip

{\tt Table I.} COVID-19 cases in China; day 1 is January 21. Source: WHO situation reports \cite{WHOrep}. Note that on Day 28 (February 17) the method of counting was changed (clinical evidence being considered sufficient even without laboratory test), leading to a sudden jump in the number of cases.

\bigskip
    \end{center}

\subsection{Benchmark: Korea}

The Republic of Korea (also known as South Korea; population around 52,000,000) is another country severely struck by COVID-19; this is also a relevant benchmark for the Italian situation given the similar population in the two countries and the similar political system (parliamentary democracy). In this case there was a sudden increase in the number of cases starting from February 19. In this case the restrictive measures also led to some reduction in the propagation speed of the epidemics, albeit of a smaller amount than in China.

In this case the ``initial'' fit refers to days from February 18 to February 24, while the ``final'' one to days from  February 24 to March 2.

In this case, the parameter $\a$ passed from $\a_i = 0.56$ to $\a_f = 0.26$; correspondingly, the doubling time passed from $\tau_i = 1.23$ to $\tau_f = 2.68$ and the daily growth factor from $\ga_i= 1.75 $ to $\ga_f = 1.29$.

\begin{center}
{\small

\begin{tabular}{|r|r||r|r||r|r||r|r||r|r||}
\hline
day & cases & day & cases & day & cases & day & cases & day & cases \\
\hline
1 &  30 & 2 & 31 & 3 & 51 & 4 & 104 & 5 & 204 \\
6 & 346 & 7 & 602 & 8 & 763 & 9 & 977 & 10 & 1261 \\
11 & 1766 & 12 & 2337 & 13 & 3150 & 14 & 3736 & 15 & 4212 \\
\hline
\end{tabular}

}

\medskip

{\tt Table II.} COVID-19 cases in the Republic of Korea; day 1 is January 17. \par Source: WHO situation reports \cite{WHOrep}.

\bigskip
    \end{center}

\begin{figure}
  \includegraphics[width=250pt]{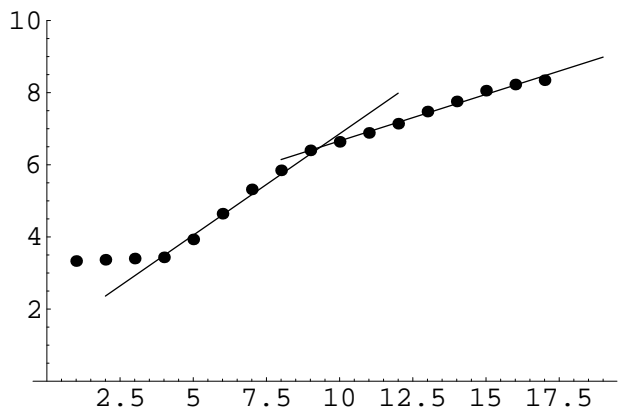}
  \\
  \caption{Epidemiological data for the Republic of  Korea, as provided by WHO \cite{WHOrep}. These are fitted by exponential laws \eqref{eq:alpha} corresponding to best fits for days 4-10 and for days 10-17.}\label{fig:korea}
\end{figure}

\subsection{National and regional data in Italy}

The statistics of data from Northern Italy is (luckily) more limited, both for what concerns the number of cases and the length of the time series. On the other hand this makes that these time series, in view of the relative large incubation period for the COVID-19 infection, did not have the time to react to the restrictive measures taken by the Italian government and to the modified habits of the population; thus in building our model we will consider the epidemic parameters as being constant over the time covered by these data.

The epidemic developed so far mainly in three regions: Lombardia, Veneto and Emilia-Romagna; from now on these will be denoted as L, V, and ER respectively (data referring to all of Italy will be denoted as I).

The total population of Italy is about 60,000,000; that of the mostly involved regions is roughly as follows: Lombardia 10,000,000; Veneto 5,000,000; Emilia-Romagna 4,500,000. The two isolated areas (red areas) have a population of 50,000 for the Lombardia one and 3,000 for the Veneto one.

Publicly available data are divided for regions (we have not analyzed so far special data about the ``red areas'') and -- limitedly to the overall count for all of Italy -- by the state of the patient; in particular these are subdivided into the following classes (identified by an acronym in the following):
Hospitalized in Intensive Care units (IC); Hospitalized in standard care units (SC); Isolated at home (HI); Recovered (REC); Dead (D).

At the date of this study, as far as we know, shortage of IC facilities was not arising, so we assume that the state of the patient is well reflected by the type of treatment he/she is undergoing.

\begin{figure}
  \includegraphics[width=250pt]{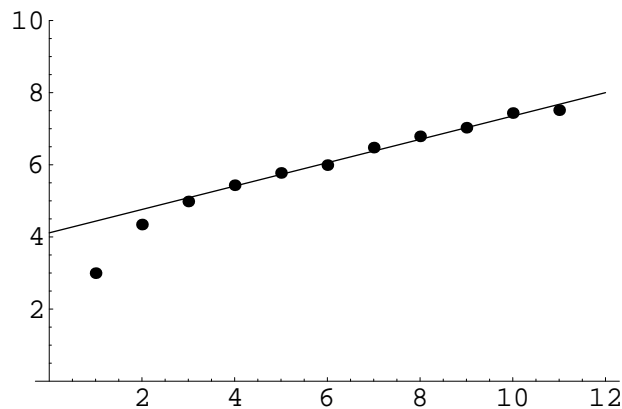}\\
  \caption{Data for the whole of Italy together with the best exponential fit for days from February 23 on}\label{fig:italy}
\end{figure}

\begin{figure}
  \includegraphics[width=200pt]{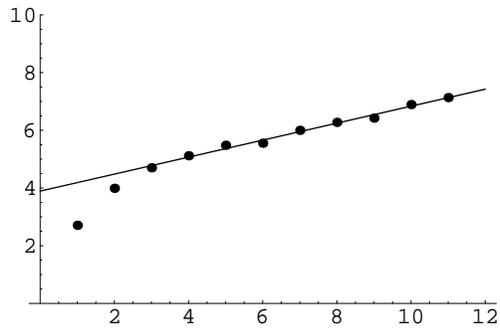}\\
  \caption{Data for the region of Lombardia together with the best exponential fit for days from February 23 on}\label{fig:lombardia}
\end{figure}

\begin{figure}
  \includegraphics[width=200pt]{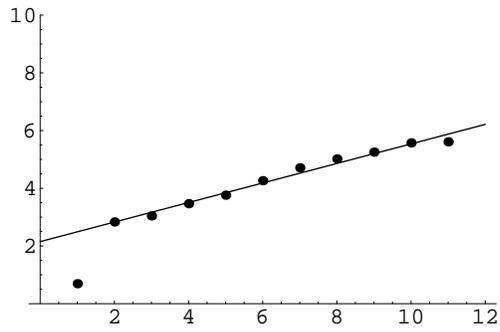}\\
  \caption{Data for the region of Veneto together with the best exponential fit for days from February 23 on}\label{fig:veneto}
\end{figure}

\begin{figure}
  \includegraphics[width=200pt]{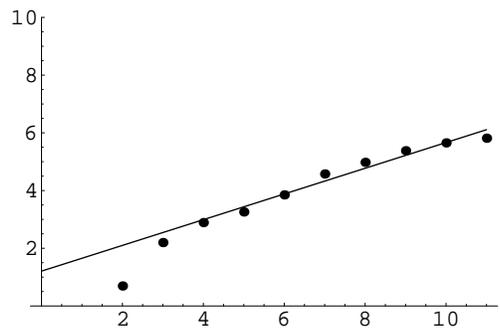}\\
  \caption{Data for the region of Emilia-Romagna together with the best exponential fit for days from February 24 on}\label{fig:emilia}
\end{figure}

The data for all of Italy and for the different regions are reported in figs.\ref{fig:italy} --\ref{fig:emilia}; these are also fitted by an exponential law, but here the fit is performed on data from day 3 (February 23) on, as in the first days there seem to be an anomalous growth, most probably due to recognizing as COVID-19 occurrences of cases which had been previous considered as standard flu with complications.\footnote{This is indeed a delicate point in the statistical analysis, in particular in the early stage of the epidemic: as the symptoms are very similar, it has to be assumed that a number of COVID cases were not recognized when it was thought the virus had not yet struck the country -- needless to say, the same could happen in analyzing data from other countries.}

The fit for the whole of Italy, and for the different considered regions, produces the following parameter $\a$ and related quantities:

\begin{center}
\begin{tabular}{|l||c|c|c||}
\hline
 & $\a$ & $\tau$ & $\ga$ \\ 
\hline
Italy     & 0.32 & 2.14 & 1.38 \\
Lombardia & 0.29 & 2.36 & 1.34 \\
Veneto    & 0.33 & 2.05 & 1.40 \\
Emilia-Romagna & 0.45 & 1.56 & 1.56 \\
\hline
\end{tabular}

\medskip

{\tt Table III.} Fit of the parameters $\a$, $\tau$, $\ga$ as defind in eqs. \eqref{eq:alpha}, \eqref{eq:tau}, \eqref{eq:gamma} for the whole of Italy and for the different considered regions.
\end{center}
\bigskip

The data available for all of Italy (I) and for these specific areas (denoted respectively as L, V and E from now on) and concerning all the recognized infected individuals, independently of the level of the symptoms, are as follows

\begin{center}
\begin{tabular}{|l||r||r|r|r||}
\hline
day & I & L & V & ER \\
\hline
21 Feb &   20 &   15 &   2 &   0 \\
22 Feb &   77 &   54 &  17 &   2 \\
23 Feb &  146 &  110 &  21 &   9 \\
24 Feb &  229 &  167 &  32 &  18 \\
25 Feb &  322 &  240 &  43 &  26 \\
26 Feb &  400 &  258 &  71 &  47 \\
27 Feb &  650 &  403 & 111 &  97 \\
28 Feb &  888 &  531 & 151 & 145 \\
29 Feb & 1128 &  615 & 191 & 217 \\
01 Mar & 1694 &  984 & 263 & 285 \\
02 Mar & 1835 & 1254 & 273 & 335 \\
\hline
\end{tabular}

\medskip

{\tt Table IV.a.} Known cases of contagion in all of Italy (I) and in different regions: Lombardia (L), Veneto (V), Emilia-Romagna (ER). Source: Presidenza del Consiglio - Dipartimento della Protezione Civile, daily press communications at 18:00 each day \cite{PCrep}. See Figure \ref{fig:italy} for a plot of these data.
\bigskip

\begin{tabular}{|l||r|r|r|r|r||r||}
\hline
day & IC & SC & HI & Rec & Dead & Total \\
\hline
24 Feb & 27 & 101 & 94 & 1 & 5 & 229 \\
25 Feb & 35 & 114 & 162 & 1 & 10 & 322 \\
26 Feb & 36 & 128 & 221 & 3 & 12 & 400 \\
27 Feb & 56 & 248 & 284 & 45 & 17 & 650 \\
28 Feb & 64 & 345 & 412 & 46 & 21 & 888 \\
29 Feb & 105 & 401 & 543 & 50 & 29 & 1128 \\
01 Mar & 140 & 639 & 798 & 83 & 34 & 1694 \\
02 Mar & 166 & 742 & 927 & 149 & 52 & 1835 \\
\hline
\end{tabular}

\medskip

{\tt Table IV.b.} Known cases of contagion in all of Italy (cumulative), according to treatment. IC: patients in Intensive Care units; SC: patients hospitalized in standard care units; HI: infected people in home isolation; Rec: recovered. Source: Presidenza del Consiglio - Dipartimento della Protezione Civile, daily press communications \cite{PCrep} and Ministero della Salute website \cite{MinSal}.
\bigskip

\begin{tabular}{|l||r|r|r||r||}
\hline
day & IC + SC & Home & Tot & Ratio \\
\hline
24 Feb & 128 &  94 & 222 & 0.42 \\
25 Feb & 149 & 162 & 311 & 0.52 \\
26 Feb & 164 & 221 & 385 & 0.57 \\
27 Feb & 304 & 284 & 588 & 0.48 \\
28 Feb & 409 & 412 & 821 & 0.50 \\
29 Feb & 506 & 543 & 1049 & 0.52 \\
01 Mar & 779 & 798 & 1577 & 0.51 \\
02 Mar & 908 & 927 & 1835 & 0.51 \\
\hline
\end{tabular}

\medskip
{\tt Table IV.c.} Known cases of contagion in all of Italy; comparison of the number of hospitalized patients versus those in home isolation. The Total in fourth column is that of IC +Hosp + Home; Ratio in the last column is that of Home/(IC + SC + Home). Elaboration on data from Table IV.b.
\bigskip

\begin{tabular}{|l||r|r|r||r||}
\hline
day & IC & SC & Tot & Ratio \\
\hline
24 Feb & 27 & 101 & 128 & 0.21 \\
25 Feb & 35 & 114 & 149 & 0.23 \\
26 Feb & 36 & 128 & 164 & 0.22 \\
27 Feb & 56 & 248 & 304 & 0.18 \\
28 Feb & 64 & 345 & 409 & 0.16 \\
29 Feb & 105 & 401 & 506 & 0.21 \\
01 Mar & 140 & 639 & 779 & 0.18 \\
\hline
\end{tabular}

\medskip

{\tt Table IV.d.} COVID-19 patients hospitalized in Italy, in IC and in SC units. The last column gives the ratio IC/(IC+SC), and shows a ratio fluctuating around 0.2, in line with findings in China \cite{CDC}. Elaboration from Table IV.b.
\bigskip

\end{center}

\subsection{More detailed local data for most affected departments}

More detailed data, at the Department (Provincia) level, are available only from February 27 on; these are reported in Table V.a for some of the more severely affected Department. These are Lodi (LO, total population about 230,000), Cremona (CR, about 360,000), Piacenza (PC, about 290,000) and Pavia (PV, about 550,000), and are located (in particular the first three) in a rather homogeneous geographical area. We also consider data for the Departments of Bergamo (BG, about 1,110,000) and Brescia (BS, about 1,260,000), which lay more to the North, of Milano (MI, about 3,200,000), and of Padova (PD, about 940,000); the latter includes the red area in Vo' Euganeo.

\begin{figure}
\begin{tabular}{|c||c|}
\hline
  \includegraphics[width=150pt]{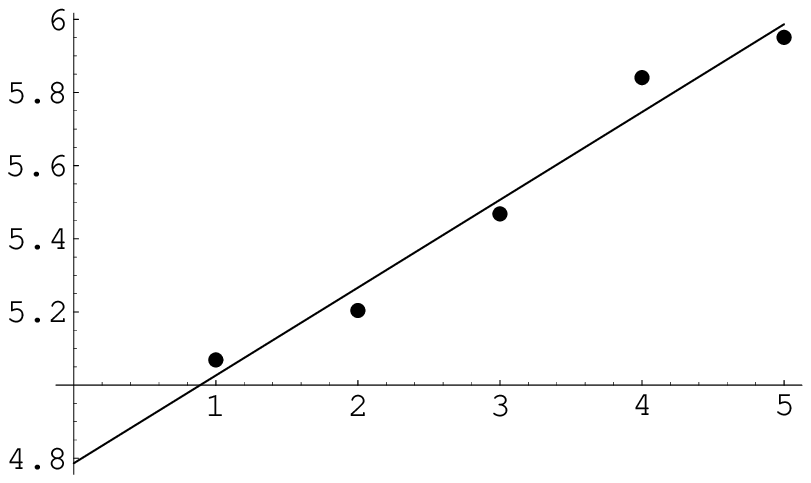} &
  \includegraphics[width=150pt]{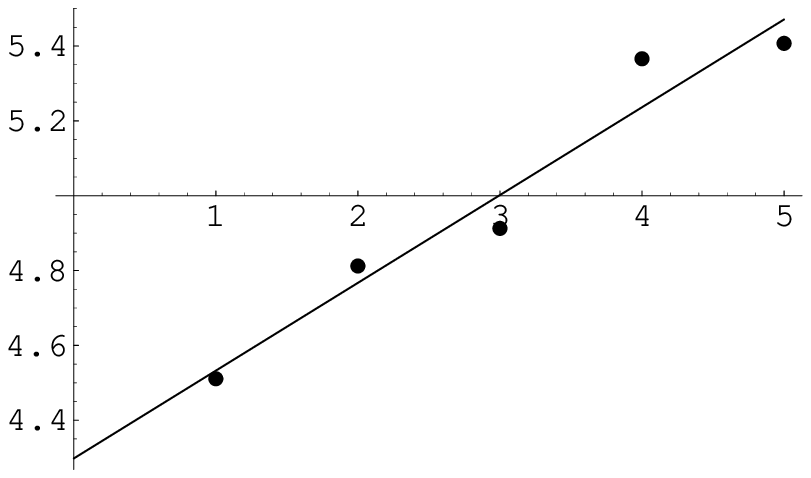} \\
  LO & CR \\
  \hline
  \includegraphics[width=150pt]{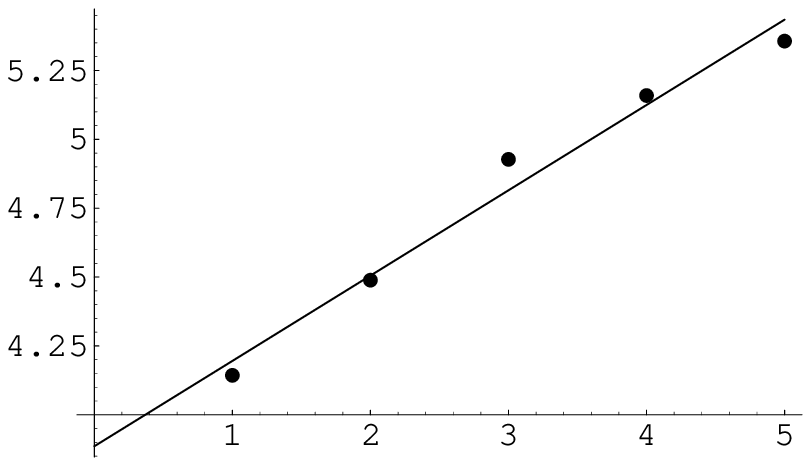} &
  \includegraphics[width=150pt]{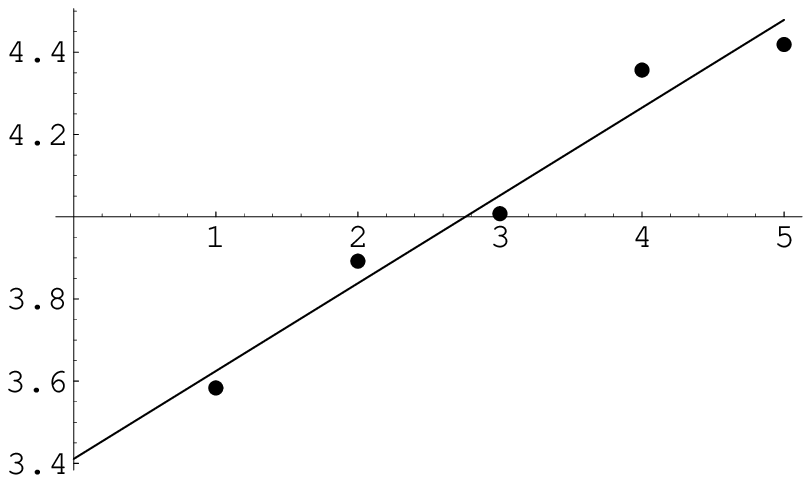} \\
  PC & PV \\
  \hline
  \includegraphics[width=150pt]{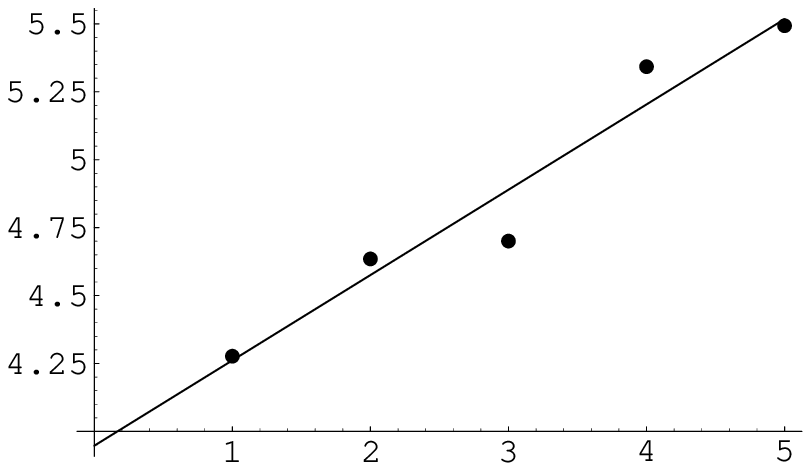} &
  \includegraphics[width=150pt]{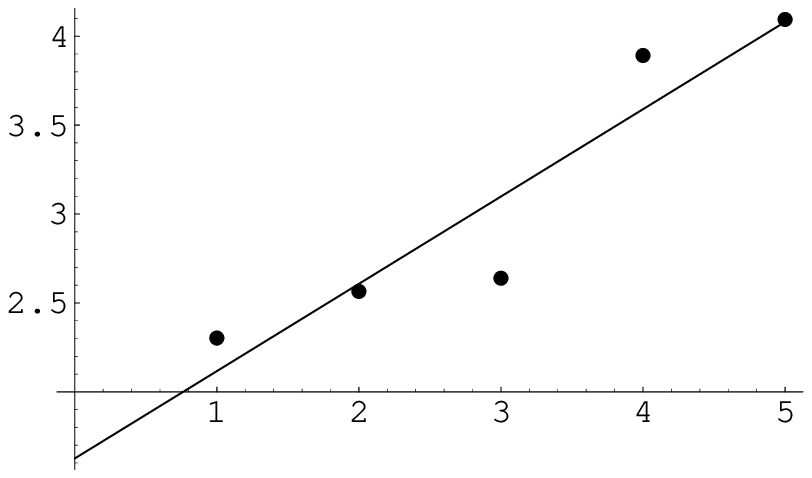} \\
  BG & BS \\
  \hline
  \includegraphics[width=150pt]{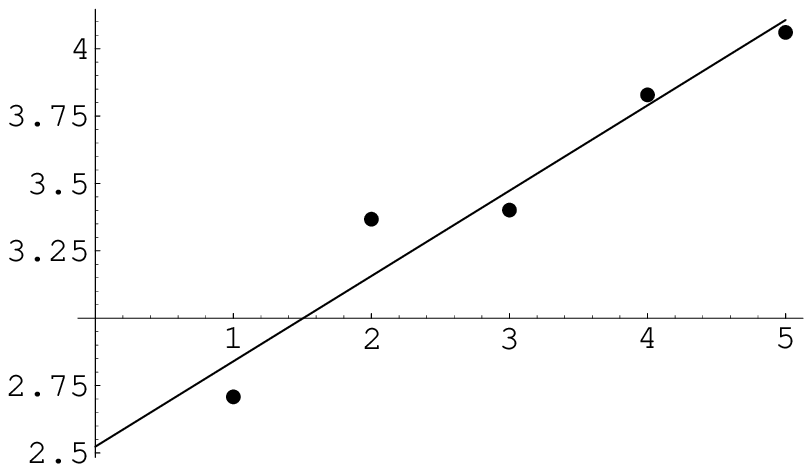} &
  \includegraphics[width=150pt]{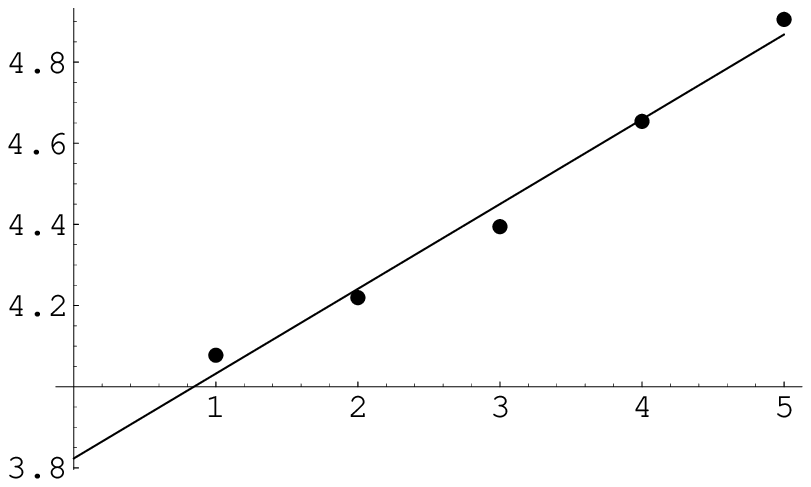} \\
  MI & PD \\
  \hline
  \end{tabular}
  \caption{Semi-logarithmic plot of data for certain Departments (see text) together with best fit (see Table V.b for value of the parameters).}\label{fig:deptgraphs}
\end{figure}

\begin{center}

\begin{tabular}{|l||r|r|r|r||r|r|r|r||}
\hline
day & LO & CR & PC & PV & BG & BS & MI & PD \\
\hline
27 Feb & 159 &  91 &  63 & 36 &  72 & 10 & 15 &  59 \\
28 Feb & 182 & 123 &  89 & 49 & 103 & 13 & 29 &  68 \\
29 Feb & 237 & 136 & 138 & 55 & 110 & 14 & 30 &  81 \\
01 Mar & 344 & 214 & 174 & 78 & 209 & 49 & 46 & 105 \\
02 Mar & 384 & 223 & 212 & 83 & 243 & 60 & 58 & 135 \\
\hline
\end{tabular}

\medskip

{\tt Table V.a.} Known cases of contagion in specific Departments including or near the Lombardia red area: Lodi (LO), Cremona (CR), Piacenza (PC) and Pavia (PV); moreover the Departments of Bergamo (BG), Brescia (BS) and Milano (MI),also in Lombardia are considered, togeher with that of Padova (PD) in Veneto, including the Veneto red area.
\bigskip

\begin{tabular}{|r||r|r|r|r||r|r|r|r||}
\hline
 & LO & CR & PC & PV & BG & BS & MI & PD \\
\hline
$\a$   & 0.24 & 0.23 & 0.31 & 0.21 & 0.31 & 0.49 & 0.32 & 0.21 \\
$\tau$ & 2.89 & 2.95 & 2.24 & 3.25 & 2.21 & 1.41 & 2.19 & 3.32 \\
$\ga$  & 1.27 & 1.26 & 1.36 & 1.24 & 1.37 & 1.63 & 1.37 & 1.23 \\
\hline
\end{tabular}

\medskip

{\tt Table V.b.} Best fit of the $\a$ factor with the corresponding doubling time $\tau$ and daily growth factor $\ga$ -- see eqs. \eqref{eq:alpha}, \eqref{eq:tau} and \eqref{eq:gamma} -- for the different Departments considered in Table V.a. See Figure \ref{fig:deptgraphs} for the fit.
\bigskip

\end{center}

We note that while Lodi, Cremona and Pavia (which is however more distant from the red area) grow at about the same rate, the growth rate in the department of Piacenza -- which is contiguous to the Lombardia red area -- appears to be specially high. The same applies to the Bergamo and Brescia. It is possible that some area in these Departments should better be isolated in the same way as the red area in Lodi Department, and indeed such measures are presently under consideration. The impact of the red area in Vo' Euganeo on the Padova Department appears to be rather limited, as also implied by its small size. The growth rate in the Milano area is higher than other ones; this had to be expected considering that it is related to the population size and density, but for exactly the same reasons it appears worrying.

\newpage

\section{Discussion}

So far we have only provided raw data; in this short Section we will present some considerations based on simple models and on projections of he progression of the epidemic with a reduced speed.\footnote{The computations in this Section are extremely rough, and will be superseeded by the more adapted model to be discussed in a companion paper; moreover they will appear rather trivial to reader expert in the field. We will nevertheless present them, as we hope this document can reach also readers not familiar with epidemic models.}

\subsection{Inference from growth rates}

The figures given above for the growth rates and daily growth factors in different areas would allow for an easy estimate of the situation in the next few days \emph{should growth go on at the same rate}. However, some restrictive measure and a campaign of public awareness were started about ten days ago, so -- in view of the COVID-19 incubation time --  these should start having some effect any moment. Comparison with the Korean experience suggest this could lead as far as halving the growth rate, i.e. doubling the replication time.

In the following Table VI we give the forecast of known infections within one week (i.e. at March 10) for different areas under different hypotheses for the reduction of the growth rate\footnote{As well known, exponential growth can not goon forever; however the numbers are still (luckily) rather small compared to the overall populations, so this regime should be well described by an  exponential behavior.}, i.e. assuming we get a reduced growth rate
\beql{eq:atilde} \widetilde{\a} \ = \ r \ \a_0 \ , \eeq with $\a$ the present one. In this Table we have considered reductions as drastic as $r =0.3$,but the Korean experience suggests it is difficult to go beyond $r=0.6 - 0.5$.

\begin{center}

{\small

\begin{tabular}{||l||r|r|r|r|r|r|r|r||}
\hline
r & 1 & 0.9 & 0.8 & 0.7 & 0.6 & 0.5 & 0.4 & 0.3 \\
 \hline
 \hline
 I & 17200 & 13800 & 11000 & 8800 &  7000 & 5600 & 4500 & 3600 \\
\hline
L & 9500 & 7800 & 6400 & 5200 & 4200 & 3500 & 2800 & 2300 \\
V & 2800 & 2200 & 1700 & 1400 & 1100 & 900 & 700 & 500 \\
ER  & 7800 & 5700 & 4200 & 3000 & 2200 & 1600 & 1200 & 900 \\
\hline
\end{tabular}

\medskip

{\tt Table VI.} Projection of infection cases at March 10 under the hypotheses the restrictive measures reduce the growth rate by a factor $r$, see \eqref{eq:atilde}.

}

\end{center}

\subsection{Suggestions from simple models}

Albeit we have not discussed models for the COVID epidemics, and we intend to present a dedicated model in a companion paper, it may be worth making some general considerations on the relevance of lowering the parameter $\a$ based on epidemiological models, and in particular on the simple SIR model \cite{Murray}.

The SIR model is described by the equations
\begin{eqnarray*}
dS/dt &=& - \ a \, S \, I \\
dI/dt &=& a \, S \, I \ - \ b \, I \\
dR/dt &=& b \,I \ .\end{eqnarray*}
Here $S$ represents Susceptible individuals, $I$ the infected ones, and $R$ those removed from the epidemic dynamics. Note that the total population $N = S+I+R$ is assumed to remain constant. 

It is immediately apparent that in the SIR model the number of infected will grow as long as $$ S \ > \ \Gamma \ := \ b/a \ ; $$ thus $\Gamma$ is also known as the \emph{epidemic threshold}. The epidemic can develop only if the population is above the epidemic threshold.

The parameters $a$ and $b$ describe the contact rate and the removal rate; they depend both on the characteristics of the pathogen and on social behavior. For example, a prompt isolation of infected individuals is reflected in raising $b$, a reduction of social contacts is reflected in lowering $\a$, and both these actions raise the epidemic threshold $\Gamma$. If this is raised above the level of the total population $N$, the epidemic stops (which means the number of infected individuals starts to decrease, albeit new individuals will still be infected). The same effect can be obtained by reducing the population $N$, i.e. by partitioning it into non-communicating compartments, each of them with a population below the epidemic threshold.\footnote{Albeit strictly speaking these predictions only hold within the SIR model, and surely the exact value of the threshold refers to this model only, the mechanism at play is rather general, and similar behaviors are met in all kind of epidemic models.}

One can easily obtain the relation between $I$ and $S$ by writing 
$$ dI/dS \ = \ - \, 1 \ + \ \frac{\Gamma}{S} \ , $$ which upon elementary integration yields
$$ I \ = \ I_0 \ + \ (S_0 \, - \, S) \ + \ \Gamma \,\log(S/S_0) \ ; $$ with $I_0,S_0$ the initial data for $I(t)$ and $S(t)$; in ordinary circumstances, i.e. unless there are naturally immune individuals, $S_0 = N - I_0 \simeq N$.

As we know that the maximum $I_*$ of $I$ will be reached when $S= \Gamma$, we obtain from this an estimate of this maximum (note that we do \emph{not} have an analytical estimate of the time needed to reach this maximum); writing $\Gamma = \s N$ (with $\s < 1$) this reads
\beq  I_* \ = \ (1 - \s ) \, N \ - \ \s N \, \log (1/\s) \ = \ \left[ 1 \ - \ \s \ - \ \s \, \log (1/\s ) \right] \ N \ . \eeq

Thus increasing $\Gamma$, even if we do not manage to take it above the population $N$, leads to a reduction of the epidemic peak; this reduction can be rather relevant also for a relatively moderate reduction of $\a$ and thus increase of $\Gamma$. See the example condensed in Table VII.

\begin{figure}
  \includegraphics[width=200pt]{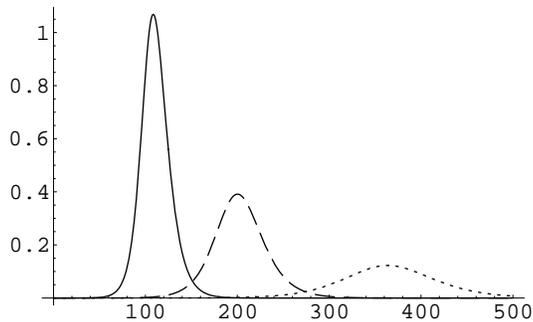}\\
  \caption{Numerical simulation of $I(t)$ as provided by a simple SIR model (in arbitrary units) for different values of the contact rate $a$; solid: $a= a_0$, dashed: $a = a_0 *0.8$, dotted: $a=a_0 *0.7$. It is clear how a reduction in the contact rate produces a substantially more manageable epidemic dynamics, both in terms of lowering of the epidemic peak and due to a slower dynamics.}\label{fig:SIR}
\end{figure}

We stress that a reduction of the parameter $a$ does not only lead to a lowering of the epidemic peak, but also slows down the whole epidemic dynamics. This is shown in Figure \ref{fig:SIR}.

\begin{center}
\begin{tabular}{|l||c|c||}
\hline
$r$ & $\Gamma$ & $I_*$ \\
\hline
1.0 & $1.21 \cdot 10^7$ & $1.81 \cdot 10^6$ \\
0.9 & $1.35 \cdot 10^7$ & $1.21 \cdot 10^6$ \\
0.8 & $1.52 \cdot 10^7$ & $6.42 \cdot 10^5$ \\
0.7 & $1.73 \cdot 10^7$ & $1.89 \cdot 10^5$ \\
0.6 & $2.02 \cdot 10^7$ & $1.02 \cdot 10^3$ \\
\hline
\end{tabular}

\medskip

{\tt Table VII.} Exemplification of the effect of reduction of the contact rate $a$ on the epidemic peak. Here $N = 20,000,000$, $b = 1/5$ and $a_0 = (5/3) 10^{-8}$,  $a = r a_0$. The values of $\Gamma = b/a$ and $I_* (\Gamma )$ are tabulated for different values of $r \le 1$ such that the population is still above the epidemic threshold, $\Gamma < N$.

\bigskip
\end{center}

We stress that the SIR model is too simple to try to extract from it any prediction on the progression of the COVID-19 epidemics; \emph{a fortiori} if with a naive estimation of the parameters. 

However, it points out at the absolute need to reduce the parameter the contact rate $a$ -- which implies reducing the parameter $\a$ describing the growth of the infection -- in order to reduce the epidemic peak \emph{and} to slow down the epidemic dynamics, getting more time to get prepared to face the epidemic peak.

\section{Conclusions}

The spread of COVID-19 epidemics in Northern Italy is well described by an exponential growth with parameter $\a \simeq 0.32$, with higher values in some regions. This should be compared with the growth rates in China, and in this sense the observed figure is quite worrying because:

\begin{enumerate}

\item The absolute value of the growth rate is rather large, and very much similar to the growth rate in the early stage of the epidemics in China ($\a = 0.33$).

\item The substantial reduction of the growth rate in China was due to rather strict measures by the central and local Governments. Restrictive measures have also been taken in Italy, but these are substantially weaker than those adopted in China. As these were implemented around February 24, their effects will be readable (in view of the rather large incubation time of COVID) only in the next few days.

\item The experience of Korea -- which is in some ways more comparable to Italy in terms of total population and political decision system -- can give some hint on the expected outcome of not so drastic restrictive measures; in that case, the $\a$ parameter has been nearly halved, and corresponding the doubling time has doubled itself. This leads -- both in Korea and assuming the same effect is obtained by the Italian restrictive measures and public awareness -- to a slowing down in the epidemic growth, but not to recovering control over the situation, i.e. taking the dynamics below the epidemic threshold \cite{Murray}.

\item We expect the growth rate of an epidemic in a given region to be proportional to, and anyway correlated to, its overall population and population density.In this respect, one should recall that the population of Hubei is around 60,000,000; on the other hand, the total population of the Italian regions mainly involved (Lombardia, Veneto, Emilia-Romagna) is about 20,000,000. This means that the virus is circulating at a very high rate, and social contacts should be substantially reduced in order to have any hope to stop, or at least substantially slowing down, its spreading.

\item The standard theory of SIR epidemics (not completely adapted to the present situation) shows that a reduction in the contact rate can lead to a substantially more manageable development of the epidemics, in particular postponing the peak period and making it less severe in terms of the number of cases to be treated simultaneously. One should not forget, indeed, that albeit COVID is lethal only in a relatively small number of cases, of the order of 1\% (to be however compared with the $10^{-4}$ of flu), this refers to the situation where patients can be properly assisted; as well known, a saturation of Hospital facilities, or even just of Intensive Care units, would lead to a dramatic increase of the mortality rate.

\end{enumerate}

\section*{Acknowledgements}

I thank Luca Peliti for useful discussions; the opinions expressed in this note are in no way involving his responsibility. I also thank Mariano Cadoni and Enrico Franco for pointing out a blur in the first version of this report. The paper was prepared over a stay at SMRI. The author is also a member of GNFM-INdAM.



\begin{thebibliography}{9}

\bibitem{PCrep}  {\tt http://www.protezionecivile.gov.it/media-comunicazione/ comunicati-stampa}

\bibitem{WHOrep} {\tt https://www.who.int/emergencies/diseases/novel-coronavirus-2019/ situation-reports}

\bibitem{CDC} The Novel Coronavirus Pneumonia Emergency Response Epidemiology Team: ``Vital Surveillances: The Epidemiological Characteristics of an Outbreak of 2019 Novel Coronavirus Diseases (COVID-19) -- China, 2020'', {\it China CDC Weekly 2020, 2(8): 113-122}.

\bibitem{ICR} {\tt http://www.imperial.ac.uk/mrc-global-infectious-disease-analysis/ news--wuhan-coronavirus/ }

\bibitem{WHOCDC} {\tt https://www.who.int/docs/default-source/coronaviruse/
    who-china-joint-mission-on-covid-19-final-report.pdf }

\bibitem{MinSal} {\tt http://www.salute.gov.it/nuovocoronavirus }

\bibitem{Murray} J.D. Murray, {\it Mathematical Biology. I: An Introduction}, Springer (Berlin) 2002

\end{thebibliography}
\end{document}